\documentclass[11pt,amsmath,amssymb,noeprint]{revtex4}
\usepackage{amsfonts}
\usepackage{amsmath}
\usepackage{amssymb}
\usepackage{array}
\usepackage{bbding}
\usepackage{bm}
\usepackage{booktabs}
\usepackage{braket}
\usepackage{breqn}
\usepackage{color}

\usepackage{dcolumn}
\usepackage{diagbox}
\usepackage{endnotes}
\usepackage{epsfig}
\usepackage{epsf}
\usepackage{epstopdf}
\usepackage{extarrows}
\usepackage{float}
\usepackage{graphicx}
\usepackage{graphics}
\usepackage{harpoon}
\usepackage{hyperref}
\usepackage{indentfirst}
\usepackage{makecell}
\usepackage{mathrsfs}
\usepackage{mathtools}
\usepackage{multirow}
\usepackage{pifont}
\usepackage{xcolor}

\usepackage{cleveref}

\newcolumntype{C}[1]{>{\centering\arraybackslash}p{#1}}
\newcommand {\sla}[1]{ #1 \!\!\!/}
\newcommand{\RM}[1]{\textrm{\uppercase\expandafter{\romannumeral#1}}}

\begin{document}

\title{$\gamma W$-exchange correction beyond the forward-angle limit in neutron $\beta$ decay}

\author{
Hui-Yun Cao$^{1}$\protect\footnotemark[1]\protect\footnotetext[1]{E-mail: caohy@hbnu.edu.cn},
Hai-Qing Zhou$^{2}$\protect\footnotemark[2]\protect\footnotetext[2]{E-mail: zhouhq@seu.edu.cn} \\
$^1$ School of Physics and Electronic Science, Hubei Normal University, Huangshi 435002, China\\
$^2$ School of Physics, Southeast University, Nanjing 211189, China}

\date{\today}

\begin{abstract}
In this work, we discuss the $\gamma W$-exchange contributions in neutron
$\beta$ decay with an elastic intermediate state, beyond the forward-angle limit (FAL). By decomposing the one-$W$-exchange and $\gamma W$-exchange amplitudes in terms of $16$ independent Pauli-spinor structures, we calculate the $\gamma W$-exchange corrections to the relevant coefficients. Our numerical results show that the relative corrections to the Fermi Born term $C_{\text{Born}}^{\text{F}}$ and the Gamow-Teller Born term $C_{\text{Born}}^{\text{GT}}$ are enhanced by about 8\% and 18\%, respectively. In particular, we find a non-zero contribution to $C_{\text{Born}}^{\text{GT}}$ from the axial-vector current, which is identically zero in the FAL. The corresponding effect on the extracted $V_{ud}$ from the neutron lifetime is also analyzed, and we find the correction to be at the $10^{-4}$ level.
\end{abstract}

\maketitle

%%%%%%%%%%%%%%%%%%%%%%%%%%%%%%%%%%%%%%%%%%%%%%%%%%%%%%%%
\section{Introduction}

Free neutron $\beta$ decay provides a clean probe to extract the  Cabibbo-Kobayashi-Maskawa (CKM) matrix element $V_{ud}$ \cite{PDG-2024}, whose precise determination is essential for testing CKM unitarity and constraining new physics \cite{The-role-of-vud-1,The-role-of-vud-2}. In recent years, a persistent $2\text{--}3\sigma$ tension between $|V_{ud}|$ values extracted from free neutron decay and those from superallowed $0^+\rightarrow 0^+$ nuclear decays has motivated theoretical improvements to the $10^{-4}$ precision level \cite{Hardy-Towner2020}. To achieve this precision, a rigorous treatment of electroweak radiative corrections (RCs) is indispensable \cite{Sirlin-PRL-2006,Seng-Particles2021}.

Specifically, $|V_{ud}|$ can be extracted from the neutron lifetime via the following relation \cite{Sirlin-PRD2019,Seng-Universe2023}
\begin{align}
|V_{ud}|^2 &= \frac{5024.7 \text{s}}{\tau_n(1+3\lambda^2)(1+\Delta_R^V)},
\end{align}
where $\text{s}$ is second, $\tau_n$ is the neutron lifetime, $\Delta_R^V$ represents the inner RC to the vector sector, and $\lambda\equiv g_A/g_V$ is the axial-to-vector coupling ratio. The vector coupling $g_V$ is protected by the conserved vector current (CVC) hypothesis \cite{CVC-hypothesis-1,CVC-hypothesis-2}, while $g_A$ undergoes strong renormalization by
\begin{align}
g_A^{\text{eff}}=g_A^{\text{QCD}}\left[1+\frac{1}{2}(\Delta_R^A-\Delta_R^V)\right],
\label{equation:gA-relation}
\end{align}
where $g_A^{\text{eff}}$ is the measured effective coupling, $g_A^{\text{QCD}}$ is the theoretical bare constant from lattice QCD calculations, and $\Delta_R^{A/V}$ represents the inner RC to the axial or the vector coupling constant \cite{gA-renormalization-1,Seng-PRD2019,Blunden-PRD2021}.

Furthermore, the analyses in Refs. \cite{Seng-Particles2021,Seng-JHEP2024} give the inner RCs as
\begin{align}
\Delta_R^V &= \Delta_{R}^{U} + 2\square_{\gamma W}^V, \nonumber\\
\Delta_R^A &= \Delta_{R}^{U} + 2\square_{\gamma W}^A + 2\square_{\text{int}}^A + \Delta_{R,\text{3pt} }^A,
\end{align}
where $\Delta_{R}^{U}$ is a universal correction independent of the decay channel \cite{DeltaR-U-1,DeltaR-U-2,DeltaR-U-3}, the axial-specific terms $\square_{\text{int}}^A $ and $\Delta_{R,\text{3pt} }^A$ account for higher-order interference and three-point correlation effects \cite{Hayen-2021,Seng-JHEP2024}, and the $\gamma W$-box contributions $\square_{\gamma W}^{V/A}$ \cite{CRegge-GH2009,CRegge-Blunden2011,Seng-PRL-2018,Seng-PRL-2020} constitute the primary source of theoretical uncertainty due to their dependence on non-perturbative hadronic dynamics. Following the dispersive framework constructed by Seng \cite{Seng-PRL-2018,Seng-PRD2019,Seng-Particles2021}, the $\gamma W$-box contributions $\square_{\gamma W}^{V/A}$ are decomposed into distinct physical regimes:
\begin{eqnarray}
\square_{\gamma W}^V &=& \frac{\alpha}{2\pi}\Big[C_{\text{NB}}+ C_{\text{B}}\Big] , \nonumber\\
\square_{\gamma W}^A &=& \frac{\alpha}{2\pi}\Big[d_1+d_2+d_B  \Big],
\label{equation:definition-of-CBorn-in-Refs}
\end{eqnarray}
with
\begin{eqnarray}
C_{\text{NB}}\equiv C_{\text{DIS}} +C_{\pi N+\text{Res}}+C_{\text{Regge}},
\end{eqnarray}
where $C_{\text{DIS}}$ and $d_1$ represent short-distance perturbative effects and structure-dependent higher-twist contributions . The terms $C_{\pi N+\text{Res}}$ and $d_2$ account for resonance and low-energy continuum effects, while $C_{\text{Regge}}$ describes intermediate-energy Regge-trajectory asymptotics. The long-distance elastic (Born) contributions, $C_{\text{B}}$ and $d_{\text{B}}$, are sensitive to the nucleon electromagnetic and axial form factors and correspond to the Fermi (F) and Gamow-Teller (GT) components in Ref. \cite{Hui-Yun-Cao-2025}  as
\begin{eqnarray}
C_{\text{B}} &\equiv& C_{\text{Born}}^{\text{F}}, \quad d_{\text{B}}\equiv  C_{\text{Born}}^{\text{GT}}.
\label{equation:definition-of-CBorn-Fermi-and-GT}
\end{eqnarray}

Conventionally, the Born contribution is evaluated in the forward-angle limit (FAL), which assumes negligible electron energy and momentum transfer \cite{Seng-Particles2021,Hayen-2021,Towner-1992}. Under this approximation, it is conventionally found that the Born correction projected onto the Fermi amplitude, $C_{\text{Born}}^{\text{F}}$, receives its nonvanishing contribution from the axial-vector part of the hadronic charged weak current, whereas the correction projected onto the Gamow-Teller amplitude, $C_{\text{Born}}^{\text{GT}}$, receives contributions from the vector and weak-magnetism parts of the hadronic charged weak current. To achieve a $10^{-4}$ precision, the validity of the FAL must be scrutinized. In this work, we present an evaluation beyond the FAL. Our study reveals a significant, non-zero contribution to $C_{\text{Born}}^{\text{GT}}$ from the axial-vector current. This contribution is identically zero in the FAL and has been neglected in previous studies.

The paper is organized as follows. In Sec.~\ref{sec:basic-formalism}, we outline the basic formula for the amplitudes, express the amplitudes in terms of $16$ Pauli-spinor structures, and present the corresponding tree-level coefficients beyond the FAL. The relations among these tree-level coefficients, the $\gamma W$-exchange coefficients, and $C_{\text{Born}}^{\text{F/GT}}$ are also discussed. In Sec.~\ref{sec:numerical-results}, we present our numerical results and compare them with those available in the literature. A detailed discussion of the numerical findings and our conclusion are given in Sec.~\ref{sec:summary}.

\section{Basic Formalism} \label{sec:basic-formalism}

\subsection{Amplitudes for one-$W$-exchange and $\gamma W$-exchange diagrams}
\begin{figure}[htbp]
\centering
\includegraphics[height=6cm]{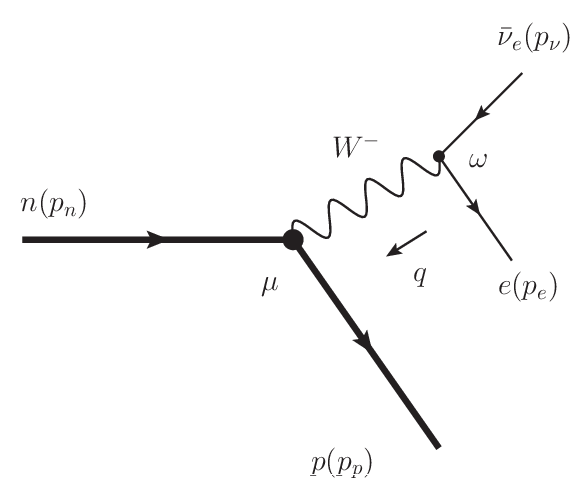}
\caption{One-$W$-exchange diagram in $n\rightarrow p e \bar{\nu}_e$.}
\label{figure:$W$-exchange-diagram}
\end{figure}

\begin{figure}[htbp]
\centering
\includegraphics[height=6cm]{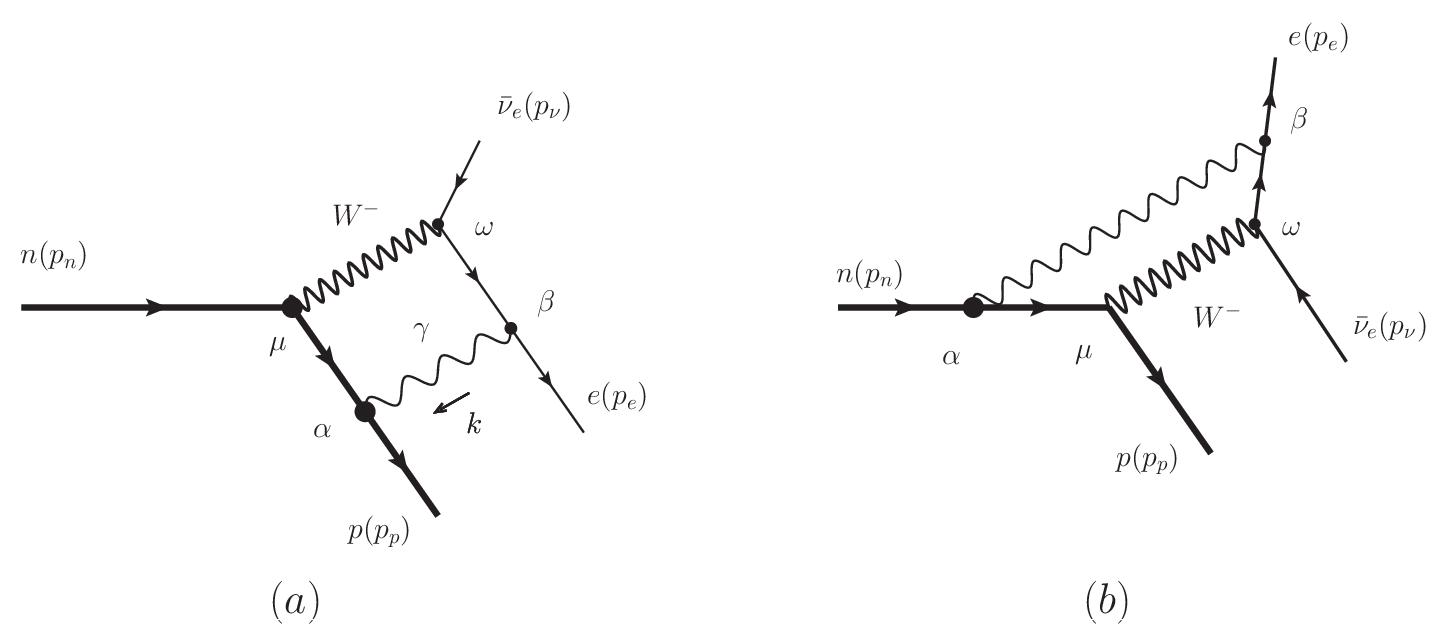}
\caption{$\gamma W$-exchange diagrams in $n\rightarrow p e \bar{\nu}_e$, where only the elastic intermediate states are considered.}
\label{figure:gamma-W-exchange-diagram}
\end{figure}

At the hadronic level, the tree diagram for free neutron $\beta $ decay is depicted in Fig.~\ref{figure:$W$-exchange-diagram}. Beyond the tree level, the $\gamma W$-exchange contributions with the nucleon intermediate state, the diagrams shown in Fig.~\ref{figure:gamma-W-exchange-diagram}, should be considered. The corresponding tree-level and $\gamma W$-exchange amplitudes can be written as
\begin{align}
\mathcal{M}^{W} &= -i\Big[\bar{u}(p_e,m_e)\Gamma^{\omega}_{W \nu e} u(p_{\nu}, m_{\nu})\Big] ~\Big[\bar{u}(p_p,m_p) \Gamma^{\mu}_{Wnp}(q) u(p_n,m_n)\Big]\frac{-ig_{\mu\omega}}{q^2-m_W^2},\notag\\
\mathcal{M}^{\gamma W}_{(a)} &= -i\int\frac{d^4 k}{(2\pi)^4}\Big[ \bar{u}(p_e,m_e) \Gamma^{\beta}_{\gamma ee}S_F(p_e+k,m_e)\Gamma^{\omega}_{W \nu e}  u(p_{\nu},m_{\nu})\Big] \notag\\
& \times\Big[\bar{u}(p_p,m_p) \Gamma^{\alpha}_{\gamma pp}(k)S_F(p_p-k,m_p) \Gamma^{\mu}_{Wnp}(q-k) u(p_n,m_n)\Big]  \frac{-ig_{\mu\omega}}{q^2-m_W^2}\frac{-ig_{\alpha\beta}}{k^2+i\epsilon},  \notag\\
\mathcal{M}^{\gamma W}_{(b)}&= -i\int\frac{d^4 k}{(2\pi)^4}\Big[\bar{u}(p_e,m_e) \Gamma^{\beta}_{\gamma ee} S_F(p_e+k,m_e)\Gamma^{\omega}_{W \nu e} u(p_{\nu},m_{\nu}) \Big] \notag\\
&\times \Big[\bar{u}(p_p,m_p)\Gamma^{\mu}_{Wnp}(q-k)S_F(p_n+k,m_n) \Gamma^{\alpha}_{\gamma nn}(k) u(p_n,m_n)\Big]  \frac{-ig_{\mu\omega}}{q^2-m_W^2}\frac{-ig_{\alpha\beta}}{k^2+i\epsilon},
\label{equation:OBE-and-TBE-amplitude}
\end{align}
where $q\equiv p_p-p_n$ is the four-momentum transfer, and $k$ denotes the momentum of the photon. The Dirac spinors $\bar{u}(p_e,m_e), u(p_{\nu},m_{\nu}),\bar{u}(p_p,m_p)$ and $ u(p_n,m_n)$ correspond to the electron, antineutrino, proton and neutron, respectively.  The propagator is given by
\begin{align}
S_F(l,m) &= \frac{i(\sla{l}+m)}{l^2-m^2+i\epsilon}.
\end{align}
The interaction vertices are defined as
\begin{align}
\Gamma^{\mu}_{\gamma ee} =& -ie\gamma^{\mu},\quad\quad\Gamma^{\mu}_{W \nu e} =i\frac{g}{2\sqrt{2}}\gamma^{\mu}(1-\gamma_5),\nonumber\\
\Gamma^{\mu}_{\gamma pp}(l) =& ie\Big[F_{1}^{p}(l^2)\gamma^{\mu}+i\frac{F_{2}^{ p}(l^2)}{2m_p}\sigma^{\mu\nu}l_{\nu}  \Big],
\nonumber\\
\Gamma^{\mu}_{\gamma nn}(l) =& ie\Big[F_{1}^{ n}(l^2)\gamma^{\mu}+i\frac{F_{2}^{ n}(l^2)}{2m_n}\sigma^{\mu\nu}l_{\nu}  \Big],\nonumber\\
\Gamma^{\mu}_{W n p}(l) =&  i\frac{g}{2\sqrt{2}}V_{ud}\Big[\Big(f_1(l^2)\gamma^{\mu}+i\frac{f_2(l^2)}{2m_N}\sigma^{\mu\rho}l_{\rho}+\frac{f_3(l^2)}{2m_N}l^{\mu}\Big)\nonumber\\
&~~~~~~~~~+\Big( f_4(l^2)\gamma^{\mu}+i\frac{f_5(l^2)}{2m_N}\sigma^{\mu\rho}l_{\rho}+\frac{f_6(l^2)}{2m_N}l^{\mu} \Big)\gamma_5\Big],
\label{equation:vertex-gammaNN}
\end{align}
where $e=-|e|$ is the electromagnetic coupling constant, $g$ is the SU(2) gauge coupling constant, and $l$ denotes the momentum of the incoming photon or $W$ boson. The average nucleon mass is defined as $m_N\equiv(m_n+m_p)/2$. The form factors (FFs) $F_{1,2}^{ p}(l^2)$ and $F_{1,2}^{n}(l^2)$ describe the proton and neutron electromagnetic structures, while $f_{1-6}(l^2)$ account for vector, weak magnetism, scalar, axial-vector, weak electricity, and induced pseudoscalar contributions, respectively.

In practical calculation, for the nucleon EM FFs, we fit them from the experimental data sets \cite{Nuclear-FFs-Exp-1,Nuclear-FFs-Exp-2,Nuclear-FFs-Exp-3,Nuclear-FFs-Exp-4,Nuclear-FFs-Exp-5,Nuclear-FFs-Exp-6,
Nuclear-FFs-Exp-7,Nuclear-FFs-Exp-8,Nuclear-FFs-Exp-9,Nuclear-FFs-Exp-10,Nuclear-FFs-Exp-11,
Nuclear-FFs-Exp-12,Nuclear-FFs-Exp-13,Nuclear-FFs-Exp-14,Nuclear-FFs-Exp-15,Nuclear-FFs-Exp-16,
Nuclear-FFs-Exp-17,Nuclear-FFs-Exp-18,Nuclear-FFs-Exp-19,Nuclear-FFs-Exp-20,Nuclear-FFs-Exp-21,Nuclear-FFs-Exp-22,
Nuclear-FFs-Exp-23,Nuclear-FFs-Exp-24,Nuclear-FFs-Exp-25} and choose the following form as
\begin{eqnarray}
F_{1}^p(l^2) &=& F_{10}^p \Big[ a_{11} G(l^2,\Lambda_{11}^2,2) + a_{12} G(l^2,\Lambda_{12}^2,2)\Big], \nonumber\\
F_{2}^p(l^2) &=& F_{20}^p \Big[ a_{21} G(l^2,\Lambda_{21}^2,3) + a_{22} G(l^2,\Lambda_{22}^2,3) \Big],  \nonumber\\
F_{1}^n(l^2) &=& F_{10}^n \Big[ a_{31} G(l^2,\Lambda_{31}^2,2) + a_{32} G(l^2,\Lambda_{32}^2,2) \Big],  \nonumber\\
F_{2}^n(l^2) &=& F_{20}^n \Big[ a_{41} G(l^2,\Lambda_{41}^2,3) + a_{42} G(l^2,\Lambda_{42}^2,3) \Big] ,
\label{equation:FFs-nuclear}
\end{eqnarray}
with $F_{10}^p=1,F_{20}^p=\mu_p-1, F_{10}^n=1, F_{20}^n=\mu_n$, $\mu_p$ and $\mu_n$ the anomalous magnetic moments of the proton and neutron, respectively. The function $G$ is defined as
\begin{eqnarray}
G(l^2,\Lambda^2,n)\equiv \frac{(-\Lambda^2)^n}{(l^2-\Lambda^2)^n}.
\label{equation:FFs-G}
\end{eqnarray}
The fitted coefficients $a_{ij}$ and cutoff parameters $\Lambda_{ij}$ are given in Sec.~\ref{sec:numerical-results}, and details of this FF parametrization can be found in Ref.~\cite{Hui-Yun-Cao-2025}.

For the weak FFs, the CVC hypothesis imposes the relations
\begin{eqnarray}
f_1 (l^2) &=&  F_{1}^p(l^2)-F_{1}^n(l^2),\nonumber\\
f_2 (l^2) &=& F_{2}^p(l^2)-F_{2}^n(l^2),
\label{equation:FFs-f1-f2}
\end{eqnarray}
and forces the scalar form factor $f_3(l^2) = 0$. In the absence of second-class currents \cite{SCC-Weinberg}, the weak electricity form factor $f_5(l^2)=0 $. The axial-vector form factor $f_4(l^2)$ is parametrized following Ref. \cite{Towner-1992} as
\begin{eqnarray}
f_4 (l^2 )&=&  g_A \frac{\Lambda_W^4}{(l^2-\Lambda_W)^2},
\label{equation:FFs-f4}
\end{eqnarray}
where $g_A$ is the axial coupling constant. The pseudoscalar FF $f_6(l^2)$ is neglected in this work, as it does not contribute to the $\gamma W$-exchange amplitude in our case.

\subsection{General form for the amplitudes and the corresponding corrections}
Following the amplitude-level framework established in Ref. \cite{Hui-Yun-Cao-2025}, we extend the calculation of the $\gamma W$-exchange contributions to kinematics beyond the FAL. Specifically, we reduce the four-dimensional amplitudes in Eq.~(\ref{equation:OBE-and-TBE-amplitude}) to a two-dimensional form and express them as a linear combination of $16$ invariant bases:
\begin{align}
\mathcal{M}^{X} &\equiv \mathcal{N}\sum_{i=1}^{16} \mathcal{C}_i^{X} O^i,
\label{equation:amplitudes-in-16-bases}
\end{align}
where
\begin{align}
\mathcal{N} \equiv -\frac{m_n\sqrt{E_\nu(E_e+m_e)}g^2V_{ud}}{4m_W^2},
\end{align}
and $X$ denotes either $W$ or $\gamma W$, $\mathcal{C}_i^X$ are the corresponding coefficients. The amplitudes $O_i$ are expressed as
\begin{align}
O_1  &\equiv \big[\xi_p^{\dagger} \xi_n \big] \big[ \xi_e^{\dagger} \bm{\sigma}\cdot\bm{n}_e \eta_{\nu}\big],
&O_2 &\equiv \big[\xi_p^{\dagger} \xi_n \big] \big[ \xi_e^{\dagger} \bm{\sigma}\cdot\bm{n}_{\nu}\eta_{\nu}\big], \notag\\
O_3  &\equiv \big[\xi_p^{\dagger} \xi_n \big] \big[ \xi_e^{\dagger} \eta_{\nu}\big],~~~~~~~~~~~~~~~~~~~~~~~~~~
&O_4 &\equiv \big[\xi_p^{\dagger} \xi_n \big] \big[ \xi_e^{\dagger} \bm{\sigma}\cdot(\bm{n}_e\times\bm{n}_{\nu})\eta_{\nu}\big], \notag\\
O_5  &\equiv \big[\xi_p^{\dagger} \bm{\sigma}\cdot\bm{n}_e \xi_n \big] \big[ \xi_e^{\dagger} \eta_{\nu}\big],
&O_6 &\equiv \big[\xi_p^{\dagger} \bm{\sigma}\cdot\bm{n}_{\nu} \xi_n \big] \big[ \xi_e^{\dagger} \eta_{\nu}\big],\notag\\
O_7  &\equiv \big[\xi_p^{\dagger} \bm{\sigma}\cdot(\bm{n}_e\times\bm{n}_{\nu}) \xi_n \big] \big[ \xi_e^{\dagger} \bm{\sigma}\cdot\bm{n}_e \eta_{\nu}\big],
&O_8 &\equiv \big[\xi_p^{\dagger} \bm{\sigma}\cdot(\bm{n}_e\times\bm{n}_{\nu}) \xi_n \big] \big[ \xi_e^{\dagger} \bm{\sigma}\cdot\bm{n}_{\nu} \eta_{\nu}\big], \notag\\
O_9  &\equiv  \big[\xi_p^{\dagger} \bm{\sigma}\xi_n \big]\times \big[ \xi_e^{\dagger} \bm{\sigma} \eta_{\nu}\big] \cdot\bm{n}_e,
&O_{10} &\equiv \big[\xi_p^{\dagger} \bm{\sigma}\xi_n \big]\times \big[ \xi_e^{\dagger} \bm{\sigma} \eta_{\nu}\big] \cdot\bm{n}_{\nu},
\notag\\
O_{11}  &\equiv \big[\xi_p^{\dagger} \bm{\sigma} \xi_n\big]\cdot \big[ \xi_e^{\dagger} \bm{\sigma} \eta_{\nu}\big],
&O_{12} &\equiv \big[\xi_p^{\dagger} \bm{\sigma}\cdot\bm{n}_e \xi_n \big] \big[ \xi_e^{\dagger} \bm{\sigma}\cdot\bm{n}_e \eta_{\nu}\big],
\notag\\
O_{13}  &\equiv \big[\xi_p^{\dagger} \bm{\sigma}\cdot\bm{n}_e \xi_n \big] \big[ \xi_e^{\dagger} \bm{\sigma}\cdot\bm{n}_{\nu} \eta_{\nu}\big],
&O_{14} &\equiv \big[\xi_p^{\dagger} \bm{\sigma}\cdot\bm{n}_{\nu} \xi_n \big] \big[ \xi_e^{\dagger} \bm{\sigma}\cdot\bm{n}_e \eta_{\nu}\big]
\notag\\
O_{15}  &\equiv \big[\xi_p^{\dagger} \bm{\sigma}\cdot\bm{n}_{\nu} \xi_n \big] \big[ \xi_e^{\dagger} \bm{\sigma}\cdot\bm{n}_{\nu} \eta_{\nu}\big],
&O_{16} &\equiv \big[\xi_p^{\dagger} \bm{\sigma}\cdot(\bm{n}_e\times\bm{n}_{\nu}) \xi_n \big] \big[ \xi_e^{\dagger}  \eta_{\nu}\big],
\end{align}
where $\xi,\eta$ are the Pauli spinors of particle and antiparticle, $\bm{\sigma}$ is the Pauli matrix, and the unit vectors $\bm{n}_e$ and $\bm{n}_{\nu}$ point along the electron and antineutrino three-momenta, respectively. The first four terms ($i=1,\dots,4$) correspond to the Fermi part of the amplitude, while the remaining terms ($i=5,\dots,16$) describe the GT part of the amplitude. Accordingly, the decomposed amplitudes are expressed as
\begin{align}
\mathcal{M}_{\text{F}}^{X} &\equiv \mathcal{N}\sum_{i=1}^4 \mathcal{C}_i^{X} O^i, \quad
\mathcal{M}_{\text{GT}}^{X}  \equiv \mathcal{N}\sum_{i=5}^{16} \mathcal{C}_i^{X} O^i.
\end{align}

In the practical calculation, we adopt the approximation $m_\nu \approx 0$ for the antineutrino mass. The four-dimensional Dirac traces are evaluated by FeynCalc \cite{FeynCalc}, the loop integrations are performed with Package-X \cite{PackageX}, and the package LoopTools \cite{LoopTools} for double-checking.

For the one-$W$-exchange contribution beyond FAL, we expand the coefficients $\mathcal{C}^{ W}_i $ in powers of $m_W^{-1}$ and $m_n^{-1}$. At leading order (LO), we obtain
\begin{eqnarray}
\mathcal{C}^{ W}_{2,\text{LO}} &=& -\mathcal{C}^{ W}_{3,\text{LO}} = g_V,\nonumber\\
\mathcal{C}^{ W}_{6,\text{LO}} &=& i\mathcal{C}^{ W}_{10,\text{LO}}= -\mathcal{C}^{ W}_{11,\text{LO}} = g_A,
\end{eqnarray}
while the remaining 11 coefficients vanish at this order. The corresponding LO amplitude for one-$W$-exchange can be written as
\begin{eqnarray}
\mathcal{M}^{W}_{\text{LO}} &=& \mathcal{N} \Big[g_V(O_{2}-O_{3}) + g_A(O_6 -i O_{10} - O_{11})\Big] .
\label{equation:$W$-exchange-amplitude}
\end{eqnarray}
In contrast, the analytical expressions for the $\gamma W$-exchange coefficients $\mathcal{C}^{\gamma W}_i $ beyond FAL are too complex to permit a simple series expansion. Instead, we calculate them directly using the numerical methods in Sec.~\ref{sec:numerical-results}.

To quantify the $\gamma W$-exchange corrections at the amplitude level, we define the dimensionless ratios $\delta_i$ as
\begin{eqnarray}
\delta_i \equiv \frac{2\pi}{\alpha}\frac{\mathrm{Re}\big[\mathcal{C}^{\gamma W}_{i}\big]}{\mathcal{C}^{W}_{i,\text{LO}}}, \qquad \text{with~} i=2,3,6,10,11.
\label{equation:delta-i}
\end{eqnarray}
where $\delta_i$ generally depend on both $E_e$ and $\bm{n}_e\cdot \bm{n}_{\nu} $, and they become kinematically independent in the FAL. The contributions for $i=1,4,5,7,8,9,12,13,14,15,16$ are not considered at present. In the FAL, the corrections reduce to the degenerate symmetry relations
\begin{eqnarray}
\delta_2^{\text{FAL}}&=&\delta_3^{\text{FAL}},\nonumber\\
\delta_6^{\text{FAL}}&=&\delta_{10}^{\text{FAL}}=\delta_{11}^{\text{FAL}}.
\end{eqnarray}
where the index $\text{FAL}$ refers to the results in the FAL.

In the practical calculation to extract the inner RCs, we retain only the terms involving the Levi-Civita tensor in the leptonic amplitude after reducing the products of Dirac matrices, following the approach of Ref.~\cite{Towner-1992} in the FAL.  Unlike in the FAL, when going beyond FAL, the coefficients $\delta_{i}$ exhibit infrared (IR) divergences. These IR divergences are canceled by the corresponding bremsstrahlung radiative corrections at the cross section level. In the following calculation, we simply take the purely IR-divergent part to be zero, i.e., we take
\begin{eqnarray}
\frac{1}{\tilde{\epsilon}_{\text{IR}}}\equiv \frac{1}{\epsilon_{\text{IR}}}-\gamma_E+\log4\pi   \rightarrow 0,
\end{eqnarray}
where $\gamma_E$ is Euler constant.

\subsection{Correction to the unpolarized differential cross section}
The unpolarized differential cross section of  neutron $\beta$ decay can be written as
\begin{eqnarray}
\frac{d^5 \sigma^{W}_{\text{un}} }{dE_e d\Omega_e d\Omega_\nu} &\approx& F(E_e) \beta \sum_{\text{helicity}} \mathcal{M}^{ W}_{\text{LO}}\mathcal{M}^{W*}_{\text{LO} }, \nonumber\\
\frac{d^5 \sigma^{\gamma W}_{\text{un}} }{dE_e d\Omega_e d\Omega_\nu} &\approx& F(E_e) \beta \sum_{\text{helicity}} 2\text{Re}\big[\mathcal{M}^{\gamma W}\mathcal{M}^{W*}_{\text{LO} }\big],
\label{equation:unpolarized-cross-section}
\end{eqnarray}
where $F(E_e)$ is the Fermi function \cite{Fermi-function}, $\beta$ is the three-body phase factor \cite{phase-factor}, and the contribution from recoil correction ${\cal M}^{W}_{\text{NLO}}$ is not considered here. Combining the one-$W$-exchange amplitude $\mathcal{M}^{W}_{\text{LO}}$ from Eq.~(\ref{equation:$W$-exchange-amplitude}), the definition of $\delta_i$ in Eq.~(\ref{equation:delta-i}), and the general expression in Eq.~(\ref{equation:unpolarized-cross-section}), we obtain
\begin{eqnarray}
\frac{d^5 \sigma_{\mathrm{un}}^{W} }{dE_e d\Omega_e d\Omega_\nu}&\approx &   8 \mathcal{N}^2 F(E_e) \beta  (g_V^2+3g_A^2), \nonumber\\
\frac{d^5 \sigma_{\mathrm{un}}^{\gamma W} }{dE_e d\Omega_e d\Omega_\nu}&\approx&  8 \mathcal{N}^2 F(E_e) \beta \frac{\alpha}{2\pi}\Big[ g_V^2  (\delta_2+\delta_3 )+g_A^2 (\delta_6+
2\delta_{10}+3\delta_{11} )\Big].
\end{eqnarray}
It is straightforward to verify that $C_{\text{Born}}^{\text{F}/\text{GT}}$ defined in Eq.~(\ref{equation:definition-of-CBorn-Fermi-and-GT}) can be expressed as
\begin{eqnarray}
C_{\text{Born}}^{\text{F}} &\equiv& \frac{1}{2} \frac{\int  dE_e d\Omega_e d\Omega_\nu F(E_e)\beta \big(\delta_2+\delta_3 \big)}{\int  dE_e d\Omega_e d\Omega_\nu F(E_e)\beta},  \nonumber\\
C_{\text{Born}}^{\text{GT}} &\equiv& \frac{1}{6} \frac{\int dE_e d\Omega_e d\Omega_\nu F(E_e)\beta \big(\delta_6+2\delta_{10}+3\delta_{11} \big) }{\int  dE_e d\Omega_e d\Omega_\nu F(E_e)\beta }.
\label{equation:CBorn-integration}
\end{eqnarray}

\section{Numerical results and discussions}\label{sec:numerical-results}
In the practical numerical calculation, we adopt the following particle masses:
\begin{eqnarray}
m_n &=&939.56542 \text{~MeV},~ m_p = 938.27209\text{~MeV}, \nonumber\\
m_e &=& 0.51100 \text{~MeV},~m_{\nu}\approx 0.
\end{eqnarray}
The parameters in the FFs of Eqs. \eqref{equation:FFs-nuclear}--\eqref{equation:FFs-f4} are taken directly from Ref.~\cite{Hui-Yun-Cao-2025}:
\begin{eqnarray}
\mu_p &=& 2.793, ~\mu_n= -1.913 ,~g_M=\mu_p-\mu_n-1=3.706, \nonumber\\
g_V &=& 1, ~g_A=-1.26,  ~\Lambda_W=1.09 \text{~GeV}, \nonumber\\
a_{11}&=&0.609, ~\Lambda_{11}=0.815\text{~GeV}, ~a_{12}=0.391, ~\Lambda_{12}=1.230\text{~GeV},\nonumber\\
a_{21}&=&0.858, ~\Lambda_{21}=0.970\text{~GeV},~a_{22}=0.142,  ~\Lambda_{22}=1.598\text{~GeV},\nonumber\\
a_{31}&=&1, ~\Lambda_{31}=1.288\text{~GeV}, ~a_{32}=-1,     ~\Lambda_{32}=1.378\text{~GeV},\nonumber\\
a_{41}&=&0.348, ~\Lambda_{41}=0.699\text{~GeV}, ~a_{42}=0.652,  ~\Lambda_{42}=1.214\text{~GeV}.
\end{eqnarray}

\subsection{The $\gamma W$-exchange contributions to the coefficients $\delta_i$}

For convenience, we separate $\delta_i$ into three components as follows:
\begin{eqnarray}
\delta_i &\equiv& \delta_{i}^{A}+ \delta_{i}^{V} + \delta_{i}^{M},
\label{eq-delta_iX}
\end{eqnarray}
where the subscripts $A$, $V$, and $M$ denote contributions from the axial-vector ($g_A$), vector ($g_V$), and magnetic ($g_M$) terms of the $\gamma W$-box exchange, respectively.

\begin{figure}[htbp]
\centering
\includegraphics[height=6.5cm]{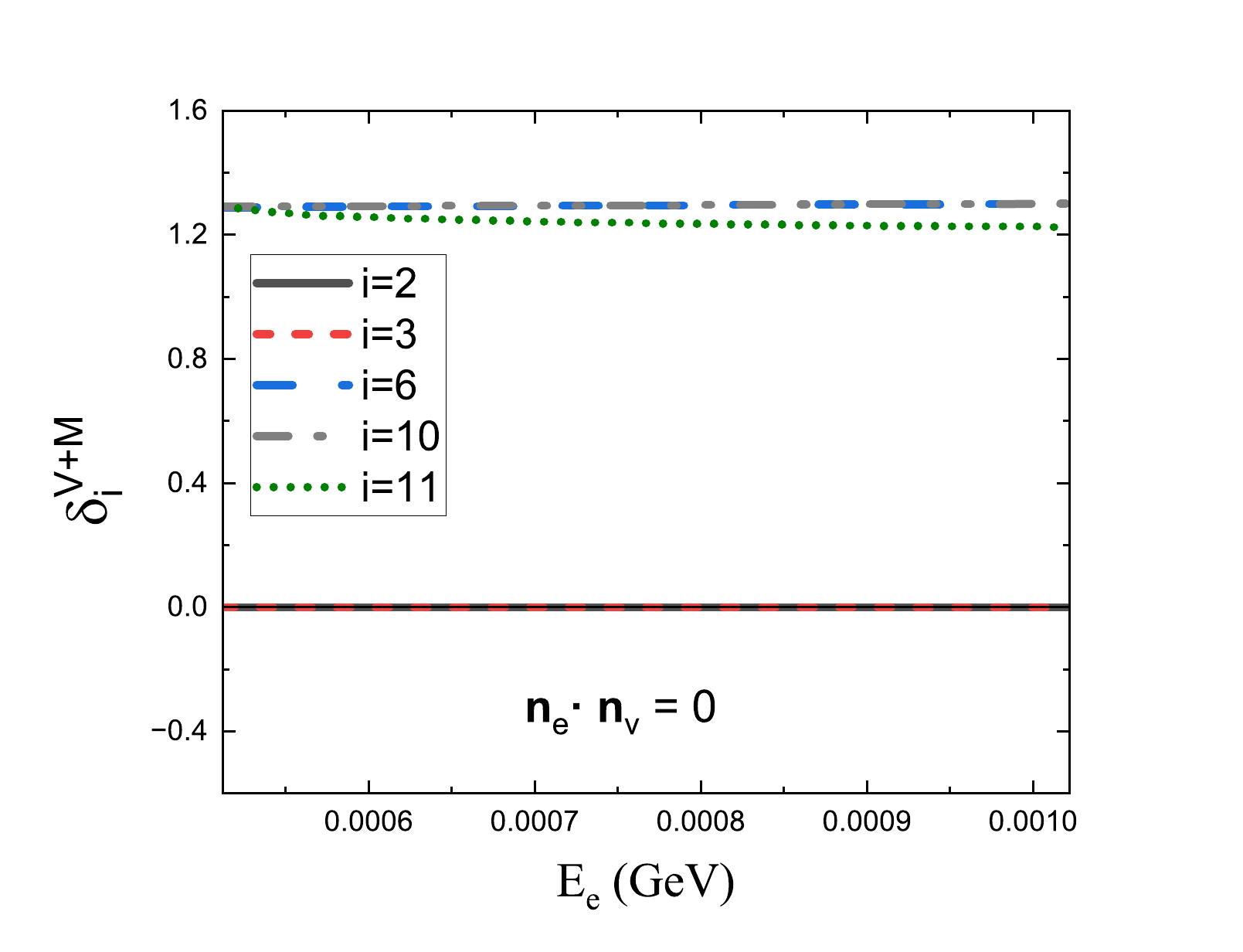}\includegraphics[height=6.5cm]{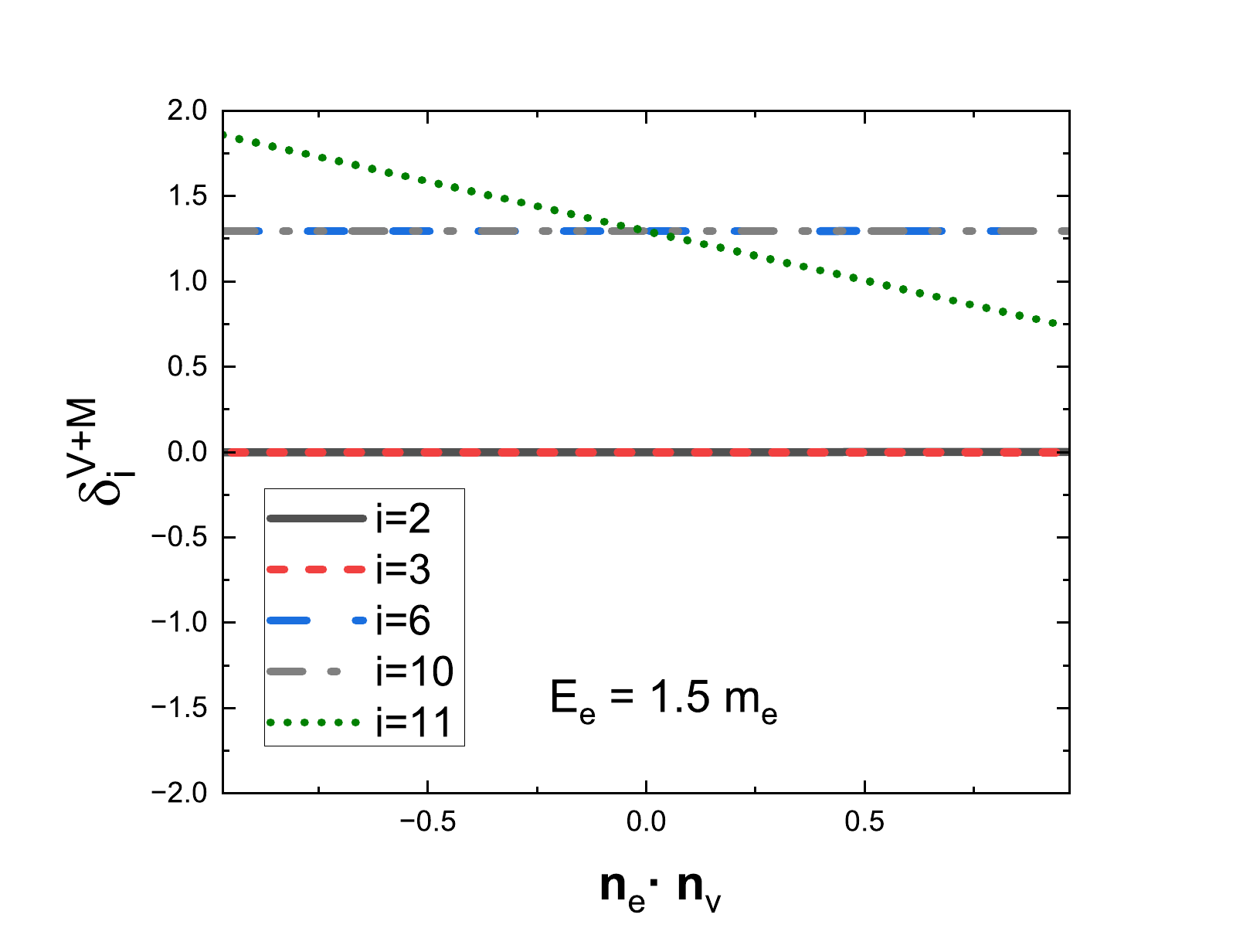}
\caption{Numerical results for $\delta_{i}^{V+M}$ with $i=2,3,6,10,11$. The left panel is the result as a function of $E_e$  at $\bm{n}_e\cdot\bm{n}_\nu=0$, while the right panel is the result as a function of $\bm{n}_e\cdot\bm{n}_\nu$ at $E_e=1.5m_e$.}
\label{figure:Re-delt-gV-gM}
\end{figure}

Fig. \ref{figure:Re-delt-gV-gM} illustrates the numerical results for the combined vector and magnetic contributions $\delta_{i}^{V+M}$ with $i=2,3,6,10,11$.  The left panel depicts the dependence on the electron energy $E_e$ at $\bm{n}_e\cdot\bm{n}_\nu=0$. All components exhibit high stability with respect to $E_e$, strongly indicating that the FAL is a very good approximation for $\delta_i^{V+M}$. The right panel presents the behavior as a function of $\bm{n}_e\cdot\bm{n}_\nu$ at $E_e=1.5m_e$. For $\delta^{V+M}_{2,3,6,10}$, the combined contributions remain essentially constant, while $\delta^{V+M}_{11}$ exhibits a distinct linear decrease as $\bm{n}_e\cdot\bm{n}_\nu$ increases from $-1$ to $1$.

\begin{figure}[htbp]
\centering
\includegraphics[height=6.5cm]{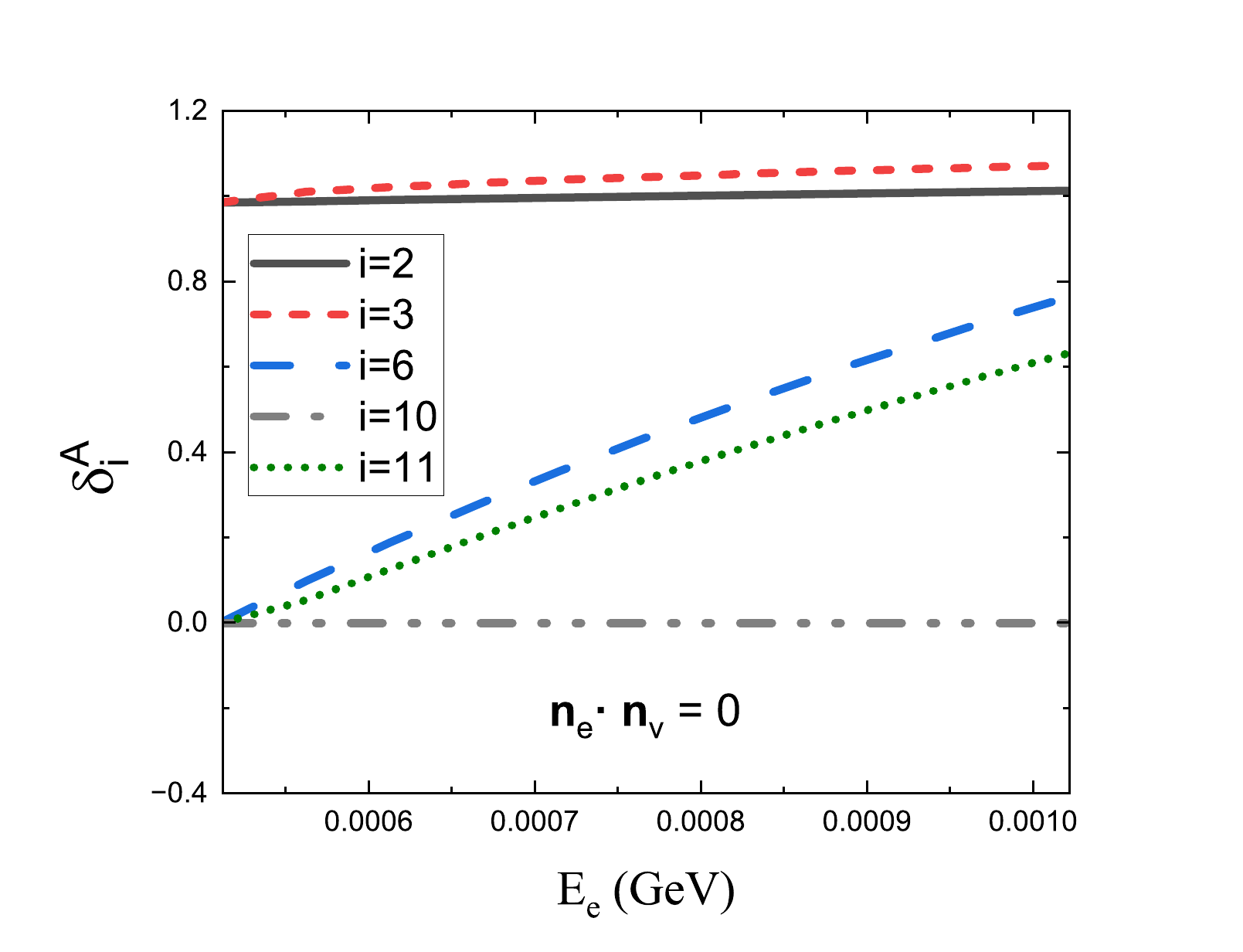}\includegraphics[height=6.5cm]{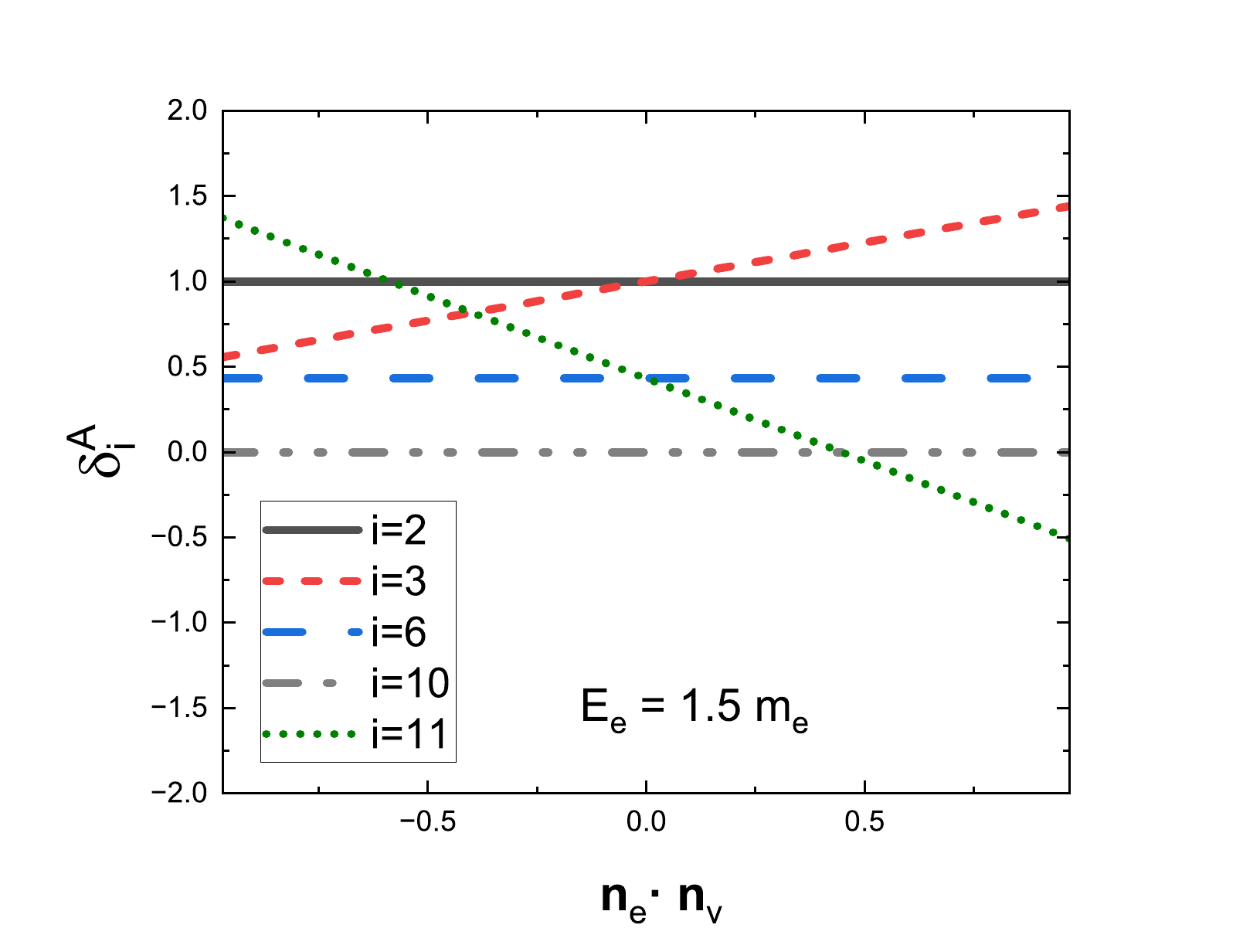}
\caption{Numerical results for $\delta_{i}^{A}$ with $i=2,3,6,10,11$. The left panel is the result as a function of $E_e$ at $\bm{n}_e\cdot\bm{n}_\nu=0$, while the right panel is the result vs. $\bm{n}_e\cdot\bm{n}_\nu$ at $E_e=1.5m_e$.}
\label{figure:Re-delta-gA}
\end{figure}

Fig. \ref{figure:Re-delta-gA} reveals the axial-vector contributions $\delta_{i}^{A}$, which exhibit markedly different behaviors. The left panel shows that $\delta^{A}_{2,3,6,11}$ evolve approximately linearly with $E_e$ at $\bm{n}_e\cdot\bm{n}_\nu=0$, while $\delta^{A}_{10}$ remains zero throughout the range. Notably, $\delta_{6,11}^{A}$ grow substantially from zero to significant magnitudes as $E_e$ increases, whereas $\delta_{2,3}^{A}$ show only slight upward trends. The right panel illustrates the angular dependence at $E_e=1.5m_e$. The terms $\delta_{2,6,10}^{A}$ are constant, whereas $\delta_{3,11}^{A}$ show strong and symmetric linear dependence on the angle.

The numerical behaviors shown in Figs. \ref{figure:Re-delt-gV-gM} and \ref{figure:Re-delta-gA} reveal that
$\delta_i^{V+M}$ and $\delta_i^{A}$ are either invariant or linearly correlated with the angular variable $\bm{n}_e\cdot\bm{n}_\nu$. Thus, for phase-space integration over the symmetric interval $[-1,1]$, we adopt the values at the symmetry point ($\bm{n}_e\cdot\bm{n}_\nu=0$) as representative averages. Furthermore, given the negligible or near-linear response to energy, the midpoint $E_e=1.5m_e$ is utilized as a reference point.
More explicitly, the angular reduction is exact for those $\delta_i$ that are independent of $\bm{n}_e\cdot\bm{n}_\nu$, and it is also exact for terms with a linear dependence on $\bm{n}_e\cdot\bm{n}_\nu$, such as $\delta_{3}^{A}$, $\delta_{11}^{A}$, and $\delta_{11}^{V+M}$, because their symmetric average over $\bm{n}_e\cdot\bm{n}_\nu\in[-1,1]$ equals the value at the symmetry point. For the energy dependence, the vector and magnetic components $\delta_i^{V+M}$ are essentially flat over the electron spectrum, so the choice of $E_e$ is immaterial for them. The axial-vector components $\delta_i^{A}$ with $i=2,3,6,11$ grow approximately linearly with $E_e$, replacing the full $E_e$ integration in Eq.~(\ref{equation:CBorn-integration}) by the midpoint value $E_e=1.5m_e$ is therefore an approximation. Since the electron spectrum is strongly weighted toward lower energies by the phase-space factor $F(E_e)\beta$, the midpoint prescription may slightly overestimate the axial-vector contribution to $C_{\text{Born}}^{\text{F/GT}}$. A direct numerical integration over $E_e$ and $\bm{n}_e\cdot\bm{n}_\nu$ with the weight $F(E_e)\beta$ would provide a more rigorous check, the present midpoint evaluation should be regarded as a transparent approximate estimate of the beyond FAL effect, sufficient for the qualitative discussion below.
The characteristic magnitudes are summarized as follows:
\begin{align}
\delta_{2}^{A}(E_e=1.5m_e,\bm{n}_e\cdot\bm{n}_\nu=0)    &\approx 0.999, &&\delta_{2}^{V+M}(E_e=1.5m_e,\bm{n}_e\cdot\bm{n}_\nu=0)     = 0, \nonumber \\
\delta_{3}^{A}(E_e=1.5m_e,\bm{n}_e\cdot\bm{n}_\nu=0)   &\approx 1.044, &&\delta_{3}^{V+M} (E_e=1.5m_e,\bm{n}_e\cdot\bm{n}_\nu=0)    = 0, \nonumber \\
\delta_{6}^{A} (E_e=1.5m_e,\bm{n}_e\cdot\bm{n}_\nu=0)   &\approx 0.433, &&\delta_{6}^{V+M}(E_e=1.5m_e,\bm{n}_e\cdot\bm{n}_\nu=0)     = 1.295, \nonumber \\
\delta_{10}^{A}(E_e=1.5m_e,\bm{n}_e\cdot\bm{n}_\nu=0)    &\approx 0,     &&\delta_{10}^{V+M} (E_e=1.5m_e,\bm{n}_e\cdot\bm{n}_\nu=0)   = 1.295, \nonumber\\
\delta_{11}^{A} (E_e=1.5m_e,\bm{n}_e\cdot\bm{n}_\nu=0)   &\approx 0.336, &&\delta_{11}^{V+M}(E_e=1.5m_e,\bm{n}_e\cdot\bm{n}_\nu=0)    = 1.238.
\label{equation:num-delta-i}\end{align}
These values effectively characterize the radiative corrections within the investigated parameter space.

\subsection{Corrections to $C_{\text{Born}}^{\text{F}}$ and $C_{\text{Born}}^{\text{GT}}$}
The numerical behaviors of $\delta_i$ in Figs. \ref{figure:Re-delt-gV-gM} and \ref{figure:Re-delta-gA} justify a simplification: given the stable $E_e$ dependence and linear symmetry over $\bm{n}_e \cdot \bm{n}_\nu$, the phase-space integration is accurately captured by evaluating the components at the kinematic midpoint ($E_e = 1.5 m_e, \, \bm{n}_e \cdot \bm{n}_\nu = 0$). Under this approximation, the integrated coefficients in Eq. (\ref{equation:CBorn-integration}) reduce to
\begin{eqnarray}
C_{\text{Born}}^{\text{F}} &=& \frac{\delta_{2}^{A}+\delta_{3}^{A}}{2}\Big|_{E_e=1.5m_e,\bm{n}_e\cdot\bm{n}_\nu=0}, \nonumber\\
C_{\text{Born}}^{\text{GT}} &=& \frac{\delta_{6}^{V+M} +\delta_{11}^{V+M}}{2}\Big|_{E_e=1.5m_e,\bm{n}_e\cdot\bm{n}_\nu=0}+\frac{\delta_{6}^{A}+3\delta_{11}^{A}}{6}\Big|_{E_e=1.5m_e,\bm{n}_e\cdot\bm{n}_\nu=0},
\end{eqnarray}
where the relation of $\delta_{6}^{V+M}=\delta_{10}^{V+M}$ has been used and the values for $\delta_i^X$ are given in Eq.~(\ref{equation:num-delta-i}).
Starting from the general definition in Eq.~(\ref{equation:CBorn-integration}), $C_{\text{Born}}^{\text{F}}$ involves the average of $(\delta_2+\delta_3)/2$ over the decay phase space. Decomposing $\delta_i=\delta_i^A+\delta_i^V+\delta_i^M$, we find numerically that $\delta_{2,3}^{V+M}=0$ at the reference kinematics which are indicated by Eq.~(\ref{equation:num-delta-i}) and Figs.~\ref{figure:Re-delt-gV-gM}--\ref{figure:Re-delta-gA}, so the vector and magnetic sectors do not contribute to $C_{\text{Born}}^{\text{F}}$ in the present calculation. The Fermi Born correction therefore comes entirely from the axial-vector sector, $\delta_{2,3}^{A}$, in agreement with the FAL result. By contrast, the GT Born correction $C_{\text{Born}}^{\text{GT}}$ receives contributions from all three sectors: the vector and magnetic parts through $\delta_{6,11}^{V+M}$, and the beyond FAL axial-vector part through $\delta_{6,11}^{A}$.

\begin{table}[htbp]
\centering
\caption{The numerical results for $C_{\text{Born}}$ in Refs. \cite{Seng-Universe2023,Hayen-2021,Hui-Yun-Cao-2025} and this work.}
\begin{tabular}{cccc}
\toprule
\hline\hline
 ~~~~~~ & ~~~~~~~~~$C_{\text{Born}}^{\text{F},A}$~~~~~~  & ~~~~~~$C_{\text{Born}}^{\text{GT},V+M}$~~~~~~ & ~~~~~~~~$C_{\text{Born}}^{\text{GT},A}$~~~~~~ \\
\midrule
\hline
Ref. \cite{Hayen-2021} &  $0.91(5)$        & $0.39(1)$ +  $0.78(2)$& 0 \\
\hline
Ref. \cite{Seng-Universe2023} &    $0.91(5)$               &    $1.22(1)$                  & 0 \\
\hline
Ref.~\cite{Hui-Yun-Cao-2025}&  $0.951 $ & $0.442  + 0.835 $ & $0$ \\
\hline
This work beyond FAL &  $1.022$ & $0.444+0.823$ & $0.240$ \\
\hline\hline
\bottomrule
\end{tabular}
\label{table:num-CBorn}
\end{table}

In Tab.~\ref{table:num-CBorn}, we compare the results beyond FAL with those in the FAL in Ref.~\cite{Hui-Yun-Cao-2025} where $C_{\text{Born}}^{\text{F}}$ and $C_{\text{Born}}^{\text{GT}}$ both as also decomposed into three parts by the indexes $A,V,M$. The Fermi Born correction from the axial-vector sector, $C_{\text{Born}}^{\text{F},A}$, increases from $0.951$ to $1.022$, corresponding to an enhancement of about $8\%$. For the GT Born correction, the vector-plus-magnetic part, $C_{\text{Born}}^{\text{GT},V+M}$, changes only slightly from $1.277(0.442  + 0.835)$ to $1.267(0.444+0.823)$. The most notable feature is a non-zero axial-vector contribution, $C_{\text{Born}}^{\text{GT},A}=0.24$, which is exactly zero in the FAL. Combining these components, the total GT Born correction increases from $1.277$ to $1.507(1.267+0.240)$, i.e., by about $18\%$, when the same precise FFs are taken as input.

For comparison, we also list the results given in Ref. \cite{Hayen-2021} and Ref. \cite{Seng-Universe2023}, which are also calculated in the FAL. We would like to mention that those results all obey some dispersion relations, which means that all the results in the FAL are consistent when the same FFs are used. In Ref.~\cite{Hui-Yun-Cao-2025}, we have discussed the results obtained with different FFs as input. In this work, we only focus on the corrections to the FAL, present the results with the most precise FFs as input, and do not discuss the dependence on the FFs.

\subsection{Correction to $|V_{ud}|$}

In this section, we discuss the corrections to $V_{ud}$ when the correction beyond the FAL is considered. For RCs which are not calculated in this work, we take them to be the same as those given in Refs.~\cite{Seng-Universe2023,Seng-JHEP2024}, where the three-point correlation term $\Delta_{R,\text{3pt}}^A$ is currently omitted pending definitive quantification from lattice QCD. The detailed RCs and the results for $V_{ud}$ are listed in Tab.~\ref{table:extracted-Vud}.
\begin{table}[htbp]
\centering
\caption{Summary of the RCs and the extracted $V_{ud}$, where the neutron lifetime $\tau_n=878.4~\mathrm{s}$ and $\lambda=-1.2754$ \cite{PDG-2024} are used. For the quantities not recalculated in this work, we use the same input values as Refs.~\cite{Seng-Universe2023,Seng-JHEP2024}.}
\begin{tabular}{c|c|c}
\toprule
\hline\hline
~~~~~~~~~~~~~~~~&~~~~~~~~Refs. \cite{Seng-Universe2023,Seng-JHEP2024}~~~~~~~~ &~~~~~~~~ This work~~~~~~~~ \\
\midrule
\hline
$\Delta_{R}^U$ & 0.01709 & 0.01709 \\
\hline
$\square_{\text{int}}^A$ & 0.00247 & 0.00247 \\
\hline
$\frac{\alpha}{2\pi}C_{\text{NB}}$ & 0.00279 & 0.00279 \\
\hline
$\frac{\alpha}{2\pi}d_1$ & 0.00248 & 0.00248 \\
\hline
$\frac{\alpha}{2\pi}d_2$ & 0.00058 & 0.00058 \\
\hline
\hline
$\frac{\alpha}{2\pi}C_{\text{Born}}^{\text{F}}$ & 0.001057 & 0.001187 \\
\hline
$\frac{\alpha}{2\pi}C_{\text{Born}}^{\text{GT}}$ & 0.001417 & 0.001751 \\
\hline
\hline
$\Delta_{R}^V$ & 0.02478 & 0.02504 \\
\hline
$\Delta_{R}^A$ & 0.02995 & 0.03062 \\
\hline
$V_{ud}$ & 0.97433 & 0.97420 \\
\hline
\bottomrule
\end{tabular}
\label{table:extracted-Vud}
\end{table}

The results in Tab.~\ref{table:extracted-Vud} show that the total GT Born correction $C_{\text{Born}}^{\text{GT}}$ increases by about $18\%$ beyond the FAL, whereas the corresponding shift in the extracted $V_{ud}$ is only about $-0.00012$, because the universal inner correction $\Delta_R^U$ provides the dominant contribution to $\Delta_R^{V/A}$. At the current precision, the beyond FAL Born correction should still be included in the analysis.
In this work, only the elastic Born contributions $C_{\text{Born}}^{\text{F/GT}}$ are evaluated beyond the FAL, the non-Born $\gamma W$ contributions are still taken from Refs.~\cite{Seng-Universe2023,Seng-JHEP2024} in the FAL. A complete assessment at the $10^{-4}$ level will eventually require extending the beyond FAL treatment to the non-Born $\gamma W$ sector, i.e., $C_{\text{NB}}$ in the vector channel and $d_1+d_2$ in the axial channel, which involve inelastic hadronic intermediate states rather than the elastic nucleon pole considered here. These inelastic $\gamma W$ components are the natural next target for a beyond FAL improvement, being more closely analogous to the kinematic relaxation studied for the Born terms in this work.

\section{Summary}\label{sec:summary}
The $\gamma W$-exchange contributions to neutron $\beta$ decay beyond the FAL are discussed. By decomposing the full amplitude into $16$ independent Pauli-spinor structures, we calculate the $\gamma W$-exchange contributions to the coefficients of the amplitudes. One of the most important and interesting properties is that a non-zero contribution to $C_{\text{Born}}^{\text{GT}}$ associated with the axial-vector coupling is found, which is exactly zero in the FAL. Together with the small shift in the vector-magnetic part, this raises the total $C_{\text{Born}}^{\text{GT}}$ by about $18\%$, while $C_{\text{Born}}^{\text{F}}$ increases by about $8\%$. The corresponding shift in the extracted $V_{ud}$ is only about $-0.00012$, because $\Delta_R^U$ dominates the total inner correction, even though the Born GT correction itself changes substantially. At the current precision, the beyond FAL Born correction should still be included. These results provide the theoretical foundation necessary for next-generation neutron decay experiments and will contribute to improved determinations of the CKM matrix element $|V_{ud}|$ and stringent tests of the Standard Model.

\section{Acknowledgments}

H.~Q.~Zhou would like to thank Zhi-Hui Guo for helpful discussions. H.~Q.~Zhou is supported by the National Natural Science Foundation of China under Grants Nos.~12150013 and 12075058. H.~Y.~Cao is supported by the Science and Technology Research Project of Hubei Provincial Education Department under Grants No.~Q20222502, Hubei Provincial Natural Science Foundation under Grants No.
2023AFB443, and by the National Natural Science Foundation of China under Grants No.~12405153.

\end{document}